\documentclass[conference]{IEEEtran}

\ifCLASSINFOpdf
\else
\fi
\usepackage{url}

\usepackage{xcolor}

\usepackage[plain]{algorithm}
\usepackage{algpseudocode}
\algnewcommand\algorithmicforeach{\textbf{for each}}
\algdef{S}[FOR]{ForEach}[1]{\algorithmicforeach\ #1\ \algorithmicdo}

\usepackage[style=ieee,backend=bibtex,bibencoding=ascii,sorting=none,maxbibnames=99]{biblatex}
\bibliography{bibliography}

\usepackage{amsmath,amssymb,amsfonts}
\usepackage{graphicx}
\usepackage{float}
\usepackage{subfig}
\usepackage{xcolor}

\newcommand{\fluo}[1]{\textcolor{black}{#1}}

\usepackage[]{fancyhdr} %
\newcommand{\changefont}{\fontsize{9}{9}\selectfont}
\IEEEoverridecommandlockouts
\begin{document}

%
\title{Optimal Connection Phase Selection of Residential Distributed Energy Resources and its Impact on Aggregated Demand}


 \author{\IEEEauthorblockN{Amina Benzerga\IEEEauthorrefmark{1},
 Alireza Bahmanyar\IEEEauthorrefmark{1},
 Damien Ernst\IEEEauthorrefmark{1}\IEEEauthorrefmark{2}
}
 \IEEEauthorblockA{\IEEEauthorrefmark{1} Department of Electrical Engineering and Computer Science,
 Liège, Belgium
 }
 \IEEEauthorblockA{\IEEEauthorrefmark{2} LTCI, Télécom Paris, 
 Institut Polytechnique de Paris,
  France}
 \IEEEauthorblockA{\{abenzerga, abahmanyar, dernst\}@uliege.be}
 \thanks{This research is supported by the public service of Wallonia within the framework of the Silver project.}
 }


%





\maketitle
\thispagestyle{fancy}
\pagestyle{fancy}


\begin{abstract}
	The  recent major increase in decentralized energy resources (DERs) such as photovoltaic (PV) panels alters the loading profile of distribution systems (DS) and impacts \fluo{higher voltage levels}. 
	Distribution system operators (DSOs) try to manage the deployment of new DERs to decrease the operational costs. However, DER location and size are factors beyond any DSO's reach.
	This paper presents a practical method to minimize the DS operational costs due to new DER deployments, through optimal selection of their connection phase.
	The impact of such distribution grid management efforts on aggregated demand for higher voltage levels is also evaluated and discussed in this paper.
	\fluo{Simulation results on a real-life Belgian network show the effectiveness of optimal connection phase selection in decreasing DS operational costs, and the considerable impact of such simple DS management efforts on the aggregated demand.}
\end{abstract}



\section*{Notation}
\begin{tabbing}
	xxxxxxxxx\= xxxxxxxxxxxxxxxxxxxxxxxxxx \kill
	\textit{Sets} \\
	$\mathcal{T}$\> Set of observation periods \\
	$\mathcal{N}$\> Set of nodes\\
	$\mathcal{E}$\> Set of edges\\
	$\mathcal{P}$\> Set of phases\\
	$\mathcal{I}$\> Set of DER\\
	\\
	\textit{Variables} \\
	$\phi_{i}$\> Connection phase of DER $i \in \mathcal{I}$ \\
	$Z_e$\> Impedance of edge $e \in \mathcal{E}$  \\
	$I_e$\> Edge current of $e \in \mathcal{E}$  \\
	$S_{n, p}$\> Power injection of $n \in \mathcal{N}$ in $p \in \mathcal{P}$ \\
	$P_{i, \phi_{i}}$\> Power injection of $i \in \mathcal{I}$ in $\phi_{i}$ \\
	$\overline{P_{i}}$\> Maximum production of DER $i \in \mathcal{I}$ \\
	$V_n$\> Voltage of $n \in \mathcal{N}$  \\
	$\overline{V}$\> Over-voltage threshold \\
	$\underline{V}$\> Under-voltage threshold \\
	
\end{tabbing}

%
\IEEEpeerreviewmaketitle

\section{Introduction}

The electrical energy production paradigm is changing.
In parallel with power plants that produce electricity in a centralized manner, the amount of decentralized electricity production is also increasing.
By installing local renewable energy production units, many traditional consumers are turning into prosumers.
\fluo{This increase in residential renewable energy sources (RES) production impacts the transmission system operation \cite{stetz2015twilight} and creates challenges resulting from the variability and uncertainty of such energy production. At distribution system (DS) level, it alters the load profile and may cause issues such as an increased phase imbalance and over-voltage.}

Phase imbalance is not a new phenomenon encountered both by distribution system operators (DSOs) and transmission system operators (TSOs). 
The imbalance usually results from an uneven allocation of loads, random consumer behaviour, and structural asymmetries \cite{ma2020review}. 
However, with the random  connection of emerging decentralized energy resources (DERs), such as Photovoltaic (PV) panels, the phase imbalance may increase considerably \cite{silva2016stochastic}. 
This, in turn, results in frequent \fluo{voltage and current issues, which are typical limiting factors for DS hosting capacity for RES \cite{torquato2018comprehensive}.
Over-voltages may lead to curtailment of DER production, which is currently only permitted to avoid network issues and to ensure power security \cite{stetz2015twilight}.}

\fluo{While DSOs might have to reinforce their network infrastructures to host substantial volumes of PV production, this represents considerable financial investment.}
There are several solutions proposed in literature to mitigate the problem \fluo{and increase the hosting capacity without reinforcing the grid} \cite{klonari2016probabilistic}. In \cite{capitanescu2014assessing, alturki2018increasing} the authors propose a network reconfiguration to maximize the hosting capacity. 
Network reconfiguration is a major DSO control tool for loss minimization and post-fault service restoration. 
However, real-time reconfiguration is not a practical solution at the LV level due to the scarcity of remotely controlled switches at this level. Active Network Management (ANM) is proposed as another solution to maximize the generation of already installed PV units through the optimal setting of available control variables \cite{linn2018enhancing,olivier2018distributed,Seydali202Photovoltaic}. 
In \cite{Hraiz2020optimal} the authors propose a method for optimal placement and sizing of PV units to reduce active power losses while achieving a high penetration level. 
However, the location and size of residential PV units are usually decided by the unit's owner based on several factors such as available area and financial means. 
There are also papers on optimal placement of equipment such as harmonic filters, battery storage units, phase-reconfiguration devices, and voltage regulators to increase the network hosting capacity \cite{Vasileios2015optimal,xu2018optimal,Sakar2017increasing,Liu2020optimal}. 
Massive integration of such devices at the LV level demands considerable investment.

This paper investigates a simple solution to manage DS operational costs by connecting new DERs, such as PV units, to optimal phases. 
The optimal selection of the connection phase of DERs is a less-investigated solution, though it is a technique which requires the least effort from the DSO's point of view. 
Such simple DS management input may have a considerable impact on higher voltage levels. 
The paper, first, presents a new method for optimal DER connection phase identification which can help the DSOs to decrease network operational costs (e.g., curtailment cost). 
Secondly, simulations are performed on a real-life Belgian network to study the impact of such local decisions on the  DS aggregated demand profile.

The \fluo{optimal phase selection} method is designed from a practical point of view. 
Each time the DSO receives a notification of the installation of a new DER by a customer, the proposed method helps to find the correct connection phase of the DER to maximize the network hosting capacity, and to minimize the operational costs due to phase imbalance during the study horizon. 
In parallel, the method checks if it is cost-effective to switch the connection phase of all DERs already installed in the network. 
Finally, the method presents the optimal decision to the DSO.

The paper is organized as follows. Section \ref{sec:methodo} presents the proposed methodology \fluo{for DER optimal connection phase selection}. Results obtained on a Belgian network case study and the discussion about the aggregated demand are shown and discussed in Section \ref{sec:casestudy}. Finally, Section \ref{sc:conclusion} concludes this paper.



\section{Optimal connection phase selection} \label{sec:methodo}

The proposed solution 
finds the most appropriate connection phase for new DERs. 
Each time a new DER needs to be added to the network, i.e., a customer buys PV panels and notifies the DSO, a first step of the algorithm finds the connection phase that leads to less over-voltage and less operational costs in a future horizon.
Then, a global optimizer finds the set of phases for each previously added DER unit that again minimize the over-voltage and operational costs. 
As rephasing existing customers with DER means DSOs will incur a labour cost, the global optimal solution is applied only when it is cost-efficient.

Consider an imbalanced three-phase distribution network.
The network topology can be represented by a tree graph $\mathcal{G} = (\mathcal{N}, \mathcal{E})$ where $\mathcal{N}$ is the set of network nodes and $\mathcal{E}$ is the set of edges linking the nodes. The set $\mathcal{P} = \{a, b, c\}$ denotes the set of phases.
The impedance of an edge $e \in \mathcal{E}$ is denoted by $Z_{e}$, representing the impedance of a three-phase cable and neutral conductor. The node voltages and edge currents are denoted by $V_{n | n \in \mathcal{N}}$ and $I_{e | e \in \mathcal{E}}$ respectively.
The network is observed over a period $\mathcal{T}$ with time steps denoted by $t$ and of length $\delta t$.
For each phase $p \in \mathcal{P}$ of node $n \in \mathcal{N}$,
the load is represented by a time-series denoted by $S_{n, p} \in \mathbb{C}^{|\mathcal{T}|}$.
The set $\mathcal{I}$ is the set of DERs. 
Let $\phi_{i} \in \mathcal{P}$ be the phase on which the DER $i \in \mathcal{I}$ is connected. 
The production time-series of a DER unit $i \in \mathcal{I}$ on phase $\phi_{i} \in \mathcal{P}$ is denoted by $P_{i, \phi_{i}} \in \mathbb{R}^{|\mathcal{T}|}$ and the maximal production of that DER unit is $\overline{P}_{i}$.

We assume that the network has an initial number of DERs and an acceptable level of imbalance. The method is designed to act as a practical tool for DSOs. 
Each time the DSO receives a request for a new DER installation at a random customer location, the proposed method performs an \fluo{individual selection (IS)} to find the optimal connection phase of the DER unit. 
The methodology inputs are DERs production $P_{i, \phi_{i}}$ and network loads $S_{n, p}$ time-series forecast for a time horizon $\mathcal{T}$. 
The optimum connection phase of the new DER is identified as the one that minimizes a cost function. The considered cost function has two main components. 
The first component is the cost directly related to imbalance. 
Several terms can be considered in the cost due to imbalance, including the equipment aging cost, additional network investment costs because of the inefficient use of its capacity, extra energy losses, and nuisance tripping. Without losing the generality of the proposed method, only the cost of network losses is considered for this study. The second component of the cost function is related to the over-voltage and it is the cost of curtailing DER production.
The cost of curtailing production and the cost due to network losses are denoted by $C^{\text{\textit{DER}}}$ and $C^{\text{\textit{NL}}}$, respectively. The minimization problem for a new DER unit $i$ is defined as follows:
\begin{subequations}
	\label{eq:maximization_problem}
	\begin{equation}
	\min_{\{\phi_{i} \}} \left( \sum_{t \in \mathcal{T}} (C^{\text{\textit{NL}}}_{t} +  C^{\text{\textit{DER}}}_{t})\right)
	\label{eq:maximization}
	\end{equation}
	subject to
	\begin{equation}
	0 \leq  P_{i, t, \phi_{i}} \leq \overline{P}_{i},\qquad \forall t \in \mathcal{T}, i \in \mathcal{I}
	\label{eq:power_range}
	\end{equation}
	\begin{equation}
	\underline{V} \leq |V_{n, t, p}| \leq \overline{V} \qquad \forall n \in \mathcal{N}, t \in \mathcal{T}, p \in \mathcal{P}
	\label{eq:voltage_range}
	\end{equation}
	where the network nodal voltages and edge currents for each time step $t$ are computed by performing a three-phase power flow: 
	\begin{equation}
	\begin{split}
	V_{n, t}, I_{e, t} = \text{\textit{\textbf{PF}}}(\mathcal{G}, P_{i, t}, S_{n, t},  Z_{e}), \\ 
	\forall n \in \mathcal{N}, e \in \mathcal{E}, t \in \mathcal{T}, i \in \mathcal{I}
	\end{split}
	\label{eq:powerflow}
	\end{equation}
	
\end{subequations}

Constraint \eqref{eq:power_range} ensures that the DER $i$ is within its production limits ($\overline{P}_{i}$).
Equation \eqref{eq:voltage_range} ensures that nodal voltages, computed through power flow \textit{\textbf{(PF)}} \eqref{eq:powerflow}, respect the voltage minimum and maximum thresholds. When voltage limits are not respected, DER productions are curtailed based on a defined policy. The curtailment ensures the acceptable level of imbalance in each time step. The pseudo code Algorithm \ref{alg:capahence} presents the IS algorithm.

\begin{algorithm}[!htb]
	\caption{\fluo{Individual selection} algorithm, where \textit{\textbf{PF}} designates the power flow computation function, \textit{cost} is the cost function and \textit{curtail} is the curtailment function.}
	\label{alg:capahence}
	\small
	\begin{algorithmic}[1]
		\Require  $\mathcal{G}$, $Z_{e | \forall e \in \mathcal{E}}$, $S_{n | \forall n \in \mathcal{N}}$, $P_{i| \forall i \in \mathcal{I}}$
		\ForEach {$\phi \in \mathcal{P}$}
		\ForEach {$t \in \mathcal{T}$}
		\State $V_{n,t}, I_{e,t} \gets \text{\textit{\textbf{PF}}}(\mathcal{G}, P_{i,t},S_{n,t},Z_e)$
		\If {$V_{n,t | \forall n \in \mathcal{N}} > \overline{V}$}
		\State $P^c_{i,t}, V_{n,t}, I_{e,t} \gets \text{curtail}(P_{i,t}, V_{n,t}, \mathcal{G},S_{n,t},Z_e) $
		\Else
		\State $P^c_{i,t}=P_{i,t}$
		\EndIf
		\State $cost_{\phi,t} += (C^{\text{\textit{NL}}}_{t} +  C^{\text{\textit{DER}}}_{t})$
		\EndFor
		\State $cost_{\phi} += \sum_{t \in \mathcal{T}} cost_{\phi,t}$
		\State $p^o=\phi^p \:| \:cost_{p}=min(cost_{\phi})$
		\EndFor
		\State \textbf{return} $p^o$
	\end{algorithmic}
\end{algorithm}

As previously mentioned, for the sake of this study, only costs caused by network losses $C^{\text{\textit{NL}}}$ and DER curtailments $C^{\text{\textit{DER}}}$ are considered.
For each time step $t$, the $C^{\text{\textit{NL}}}_t$ is defined as:
\begin{equation}
C^{\text{\textit{NL}}}_t = P^{\text{\textit{NL}}}_t \times \gamma  \times e_{price,t},
\end{equation}
where $\gamma$ is the power-to-energy conversion coefficient considering a constant power during period $\delta t$, $e_{price,t}$ is the electricity price and $P^{\text{\textit{NL}}}_t$ is the total network active power loss defined as:
\begin{equation}
P^{\text{\textit{NL}}}_t = P_{input,t} - P_{load,t},
\end{equation}
where $P_{input,t}$ is the power fed to the network:
\begin{equation}
P_{input,t} = P_{root,t} + \sum_{\forall i \in \mathcal{I}} P_{i,t}^c,
\end{equation}
where $P_{root,t}$ is the active power fed to the feeder at the root node, and is calculated from the power flow results.

The curtailments' $C^{\text{\textit{DER}}}_t$ cost is defined as:
\begin{equation}
C^{\text{\textit{DER}}}_t = \sum_{\forall i \in \mathcal{I}} (P_{i,t}-P_{i,t}^c) \times \gamma \times e_{price},
\end{equation}
where $P_{i,t}^c$ is the production of DER unit $i$ for time interval $t$ after curtailment. 
\fluo{The curtailment strategy is not a part of the methodology. 
Rather, it serves as a way to take into account over-voltage in the cost. 
Therefore, the curtailment policy is not discussed in this section.}

\begin{figure}[!htb]
	\centering
	\includegraphics[width=0.5\linewidth]{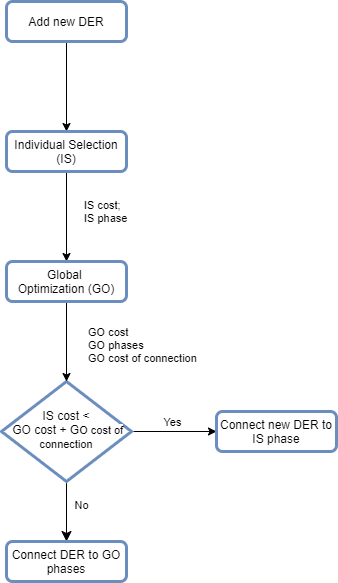}
	\caption{General architecture of the proposed method. \label{fig:method}}
\end{figure}

After installing several DER units in the network, it might reach a condition where modifying the connection phase of all previously installed DERs is cost-efficient. 
To check this, after the IS, the method runs a global optimization (GO) to find the optimal connection phase of all the DER units connected to the network. 
The optimal set of phases $\phi_{i | \forall i \in \mathcal{I}}$ is the one which minimizes the cost function:
\begin{equation}
\min_{\{\phi_{i | \forall i \in \mathcal{I}} \}} \left( \sum_{t \in \mathcal{T}} (C^{\text{\textit{NL}}}_{t} +  C^{\text{\textit{DER}}}_{t})\right)
\label{eq:Gmaximization}
\end{equation}
Global optimization is performed using a genetic algorithm.
Of course, modifying the connection phase of the all previously installed DERs has a cost. 
This is the cost to change some prosumer connection phases to coincide with the GO result.
Considering this cost aims to avoid frequent changes of connection phases.
If the IS cost is greater than the sum of the GO cost and phase-switching cost, the customers connection phase will be switched according to the GO result. 
Otherwise, the new DER unit will be connected to the optimum phase selected by the IS. A flowchart of the general architecture is shown in Fig. \ref{fig:method}.

\section{Numerical Study} \label{sec:casestudy}
The simulation study was conducted on an LV distribution network inspired by a real-life European network \cite{benzerga2021probabilistic}. The network has 128 customers and 256 nodes. 

For each new DER installation, the considered time horizon $\mathcal{T}$ is one year.
\fluo{The number of PVs in one unit is set to 13 as it is the minimal number to accommodate Belgian household consumption and the watt peak of each PV is 290W \cite{gysel_2018}.
The PV production time-series were retrieved from \cite{elia_2020}.}
The cost for switching the connection phase of a PV unit owner was arbitrarily set to 100 euros. 
The price of electricity was set to 0.2702 euros per kWh, in line with average price in Belgium in 2020 \cite{electricityPriceStatistics_2021}. 
The curtailment strategy of GO is the same as that which is considered for IS. 
The GO is carried out using the genetic algorithm package \textit{geneticalgorithm} \cite{bozorg2017meta}.

\begin{table*}[htb!]
	\centering
	\caption{The obtained costs for all new PV units. Costs are in euros. The selected option is in bold. The total GO is the sum of the labour cost (number of phase to rephase multiplied by 100€) and the GO cost column.}
	\begin{tabular}{|c|c|c|c|c|c|}
		\hline
		\begin{tabular}[c]{@{}l@{}}Total number of \\ PVs \end{tabular} &  \begin{tabular}[c]{@{}l@{}}IS cost\\phase A\end{tabular} & \begin{tabular}[c]{@{}l@{}}IS cost\\phase B\end{tabular} &
		\begin{tabular}[c]{@{}l@{}}IS cost\\phase C\end{tabular} &
		\begin{tabular}[c]{@{}l@{}}GO cost\end{tabular} &
		\begin{tabular}[c]{@{}l@{}}Total GO cost  (number of \\ customers to rephase)\end{tabular}\\ \hline
		
		46 & 22874 & 23286 & 19064 & 283 & \textbf{2983} (27) \\ \hline
		47 & 277 & \textbf{275} & 276 & 273 & 3973 (37) \\ \hline
		48 & 282 & 282 & \textbf{280} & 279 & 2779 (25) \\ \hline
		49 & \textbf{289} & 291 & 290 & 286 & 3486 (32) \\ \hline
		50 & 293 & 295 & \textbf{292} & 287 & 4087 (38)  \\ \hline
		
	\end{tabular}
	\label{tab:all_results}
\end{table*}

A simple curtailment policy, that curtails all active PV units when an over-voltage occurs, was considered for the sake of simplicity. 
\fluo{The over-voltage threshold was set to +5\% of the nominal voltage.}
More sophisticated policies such as the one proposed in \cite{olivier2018distributed} can be incorporated in the proposed architecture.

\begin{figure}[!htb]
	\centering
	\includegraphics[width=1\linewidth]{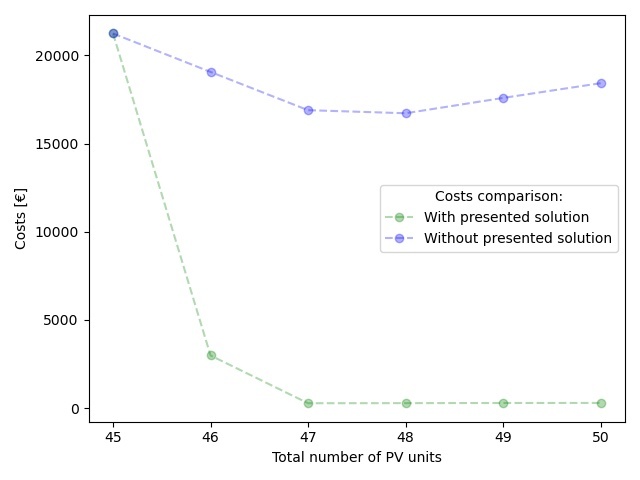}
	\caption{Comparison of costs with and without the presented solution. \label{fig:costs}}
\end{figure}

A scenario was designed to test the performance of the proposed method and to show the impact of the DSO decisions on higher voltage levels. 
In this scenario, 45 customers (i.e., 35\% of all the customers) have already installed PVs. 
The 45 customers were chosen randomly to install PV at their connection phase.
Then, the addition of \fluo{5} new PVs is considered sequentially.
The cost results for this scenario are shown in Table \ref{tab:all_results}. 
In this table, 'IS cost' for each phase refers to the cost of selecting that phase for the new PV unit without changing the phases of the previously installed PVs.
Then, 'GO cost' shows the minimum cost obtained with the optimal phases while 'total GO cost' refers 'GO cost' plus the labour cost for modifying the connection phase of previously installed PVs.
As can be seen, for the first added PV, 
the total GO cost is smaller than all IS costs.
Thus, the decision is to modify the connection phase of 27 PV units according to the GO outputted phases. 
This rephasing enables to considerably decrease the next costs as shown in Fig. \ref{fig:costs}. 
This figure depicts the difference between using the proposed solution, and connecting PVs to the customer's current connection phase. 
The cost of curtailment dominates the total cost for the first PV unit.
The costs with our solution are the ones in bold in Table \ref{tab:all_results}.
With our solution and by modifying the connection phases of PV units, the total cost is considerably decreased and is limited to the cost of network losses.

\begin{figure}[!htb]
	\centering
	\includegraphics[width=1\linewidth]{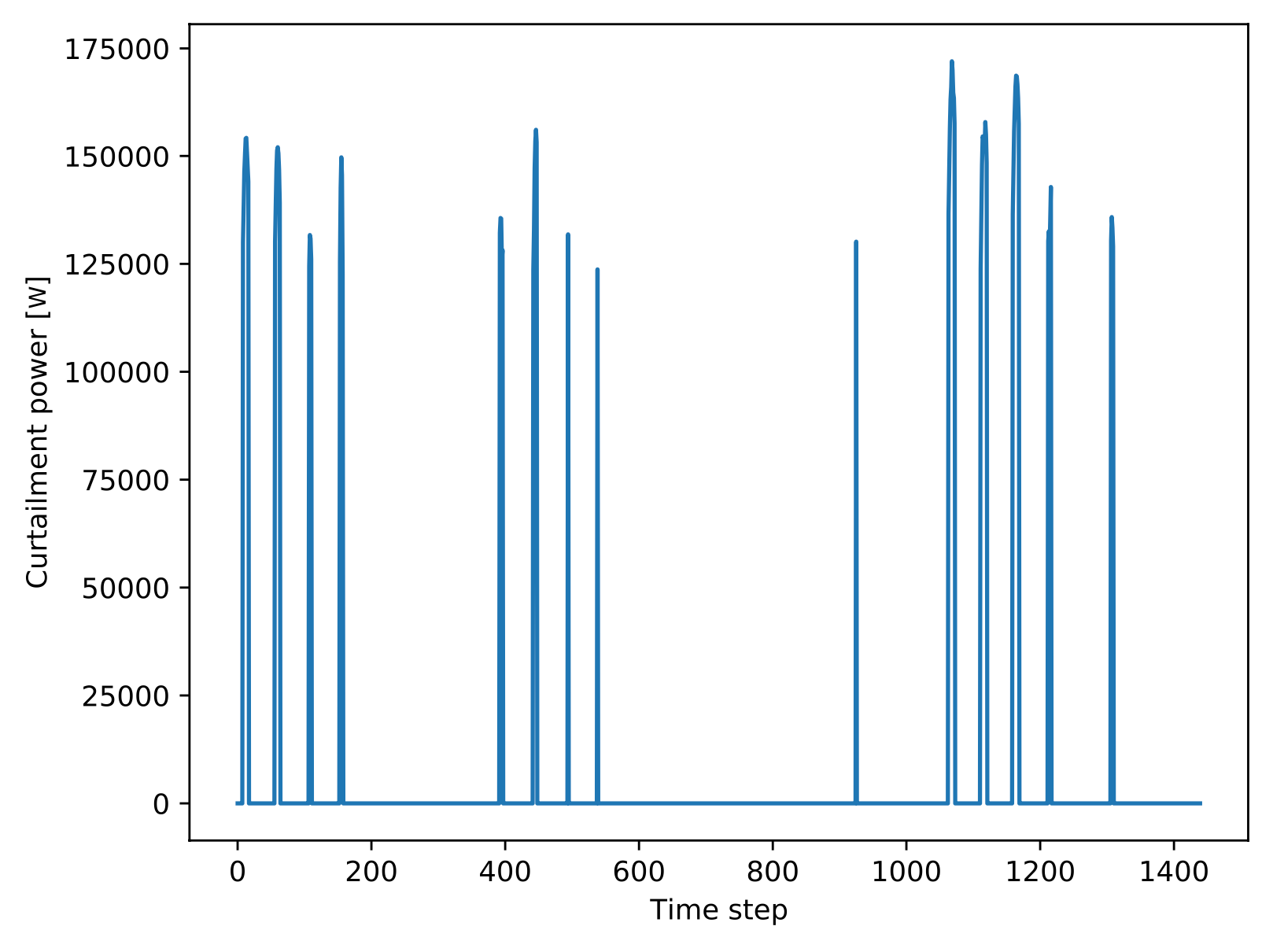}
	\caption{Simulation Curtailment powers for 46 PV units with GO. \label{fig:curtailment_scen1}}
\end{figure}

Fig. \ref{fig:curtailment_scen1} shows the total curtailment without applying the method proposed rephasing action after the first added PV unit. 
By rephasing 27 PVs, the over-voltages and thus the frequent curtailments are avoided. 

\begin{figure}[!htb]
	\centering
	\includegraphics[width=1\linewidth]{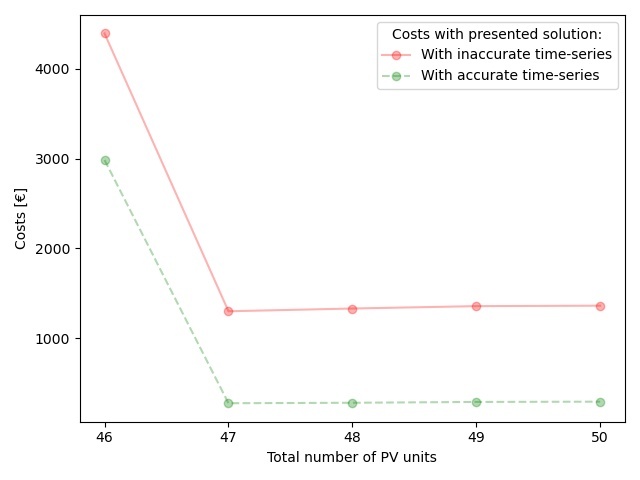}
	\caption{Comparison of costs with presented  solution using inaccurate or accurate time-series as input. \label{fig:costs_presented_solution}}
\end{figure}

Results shown in Fig. \ref{fig:costs} and Table \ref{tab:all_results} were obtained using accurate time-series for both production and consumption. 
However, in practice, such accurate time-series are not available in advance for DSOs. 
There are several studies, such as \cite{dong2019hybrid, son2020analysis, ding2021novel}, on forecasting the consumption and PV production time-series. 
To assess the performance of the proposed method without accurate time-series, the method was also evaluated using inaccurate time-series.
The load time-series were generated by random variation of each load time-series within a 30\% deviation 
with normal distribution.
The same process generated the PV production time-series within a 5\% deviation with normal distribution.
Fig. \ref{fig:costs_presented_solution} shows the costs for adding the same PVs as previously but with inaccurate time-series and accurate time-series both using the presented method.
For the first added PV, the presented method with inaccurate time-series chooses to rephase 31 customers, while 27 customers were rephrased with accurate time-series. 
For the next added PVs, the curve trend is similar to the case with accurate time-series.
Both costs of this figure are still considerably smaller than the cost without the presented method in Fig. 3.

\begin{figure}[!htb]
	\centering
	\includegraphics[width=1\linewidth]{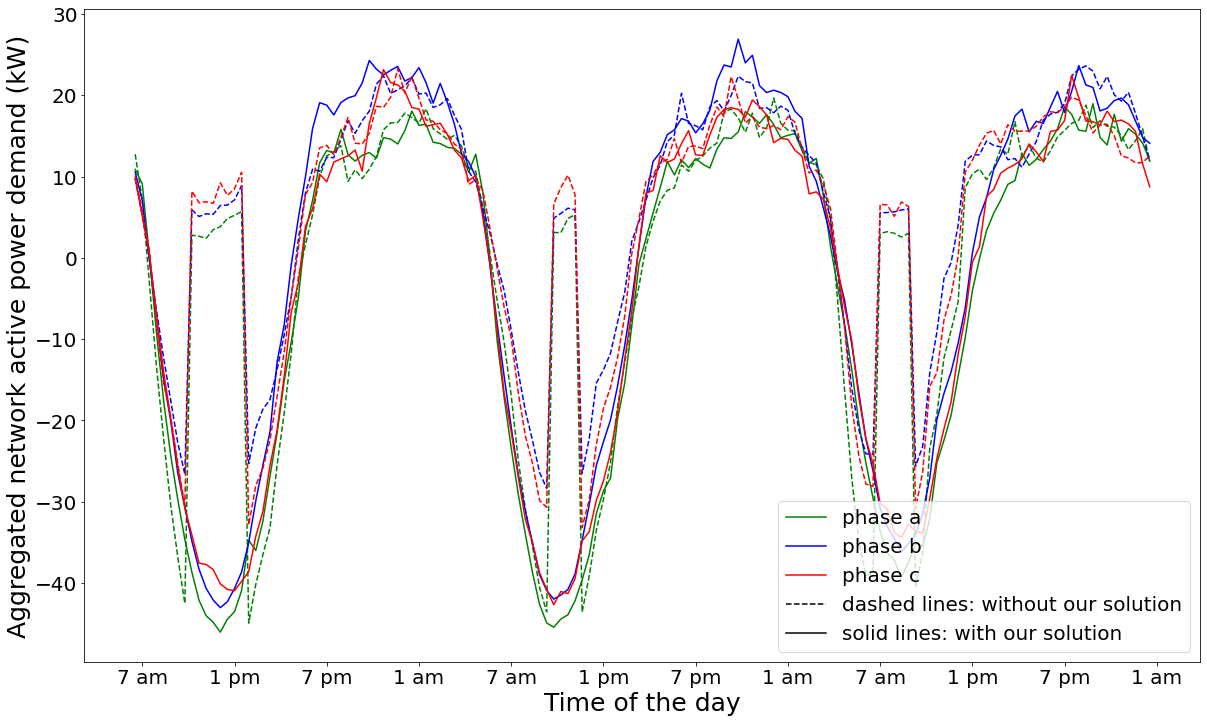}
	\caption{Aggregated active power demand by time of the day. \label{fig:input_power_scen1}}
\end{figure}

Fig. \ref{fig:input_power_scen1} shows the effect on the aggregated active power demand.
The main impact is that by avoiding frequent PV curtailments, the profile of network demand, particularly during the day, is considerably improved and phases are more balanced. 
It is interesting to note that the absolute values of demand without the presented solution is lower. This can lead to lower network losses in this case. However, since the cost of curtailment dominates the total cost, our solution maximizes the production as shown by the negative values mid-day in the figure.
This shows that simple actions taken by DSOs on LV networks can have a considerable impact on higher voltage levels.

\section{Discussion and future work} \label{sc:discussion}
The current DSO practice for new PV units is to connect them to the current customer connection phase. 
This practice may increase the network operational costs due to imbalanced operation, and may also increase the frequency of network issues such as over-voltage. 
The imbalance can be improved by proper addition of PV units to an optimal phase, but the question is how to choose the optimal phase each time the DSO receives a request for a new PV installation. 
The other question is what to do if the operational costs for all three connection phases are unacceptable. 
This paper presents a simple and practical method designed from the DSO's point of view to provide suggestions for such questions. 
The results indicate that despite the simple procedure, the method can provide good suggestions to considerably reduce the network operational costs.

The proposed method requires a forecast of PV and load time-series to consider the variation of consumption and production over time. 
There are methods to predict consumption and production profiles with a fair accuracy \cite{dong2019hybrid, son2020analysis, ding2021novel}. 
The obtained results indicate that the inaccuracies in the forecasted time-series can impact the results of the proposed method. 
However, the costs with selected phases with inaccurate inputs are still close to the optimal solution and considerably lower than those observed with the current DSO practice.

Considering both stochastic and time variations in an optimization procedure that repeats the computations over and over may not be computationally tractable. Designing a computationally efficient algorithm to consider stochasticity can be a line for future work. 
Further studies can investigate the study horizon $\mathcal{T}$, electricity price, and phase-switching cost, as they impact the final decision. 
Moreover, more cost terms could be considered as well as a more sophisticated curtailment policy.

\section{Conclusion} \label{sc:conclusion}
With massive integration of DERs, DSOs are facing increasing challenges such as greater imbalance and frequent curtailments. 
Optimal selection of the connection phase of PV units requires least effort but is a less-investigated solution to this problem. 
This paper has presented a two-step method to this end. 
The method performs an optimal selection of the connection phase of each new DER, and it simultaneously checks if it is cost-effective to modify the connection phase of all the installed network DERs to the optimal phases identified. 

Results on an LV network simulation with PV as a DER shows that using optimal selection of the connection phase helps to considerably decrease the curtailment and the over-voltage, thus the DS operational costs. 
The results also highlighted the impact of  LV management with the proposed actions on the aggregated demand. Both the imbalance and the active power demand decrease with the proposed method.







%
\section*{References}
\printbibliography[heading=none]
\end{document}